\documentclass[11pt]{article}

\usepackage{psfig,epsf}
\usepackage{graphicx}

\newcommand{\be}{\begin{equation}}
\newcommand{\ee}{\end{equation}}
\begin{document}
\title{Non-extensive Trends in the Size Distribution \\ of Coding and Non-coding DNA 
Sequences \\ in the Human Genome}
\author{Th. Oikonomou$^{1,2}$\thanks{E-mail: thoikonomou@chem.demokritos.gr}
 and A. Provata$^1$
\thanks{Corresponding author, e-mail:aprovata@limnos.chem.demokritos.gr} \\ \\
$^1$Institute of Physical Chemistry,\\
National Center for Scientific Research ``Demokritos'' \\
15310, Athens, Greece \\ \\ 
$^2$ School of Medicine, Department of Biological Chemistry \\
University of Athens,
Goudi, 11527, Athens, Greece}
\maketitle

\begin{abstract}
We study the primary DNA structure of four of the most completely
sequenced human chromosomes (including chromosome 19 which 
is the most dense in coding), using Non-extensive Statistics. We show that
the exponents governing the decay of the coding size distributions vary between
$5.2 \le r \le 5.7$ for the short scales and $1.45 \le q \le 1.50$
 for the large scales. On the contrary,
the exponents governing the decay of the non-coding size distributions 
in these four chromosomes, take the values
$2.4 \le r \le 3.2$ for the short scales and $1.50 \le q \le 1.72$ 
for the large scales.
 This quantitative difference,
in particular in the tail exponent $q$, indicates that the non-coding (coding) size 
distributions have long (short) range correlations. This non-trivial difference
in the DNA statistics is attributed to the non-conservative (conservative)
evolution dynamics acting on the non-coding (coding) DNA sequences. 
\end{abstract}



\section {Introduction}
\label{sec:intro}
\par During recent years numerous studies on the statistics of
genomic sequences have demonstrated various degrees of complexity 
in the primary structure of DNA. In particular,
Peng et al. in 1992  demonstrated the existence of
long range correlations using the ``DNA walk'' model \cite{stanley:1992}.
Similar conclusions were reached by Li et al. \cite{li:1992} and
Voss \cite{voss:1992} using the $1/f$
spectrum and later by studies on the size distribution
 of Purine (Adenine, Guanine) and Pyrimidine (Thymine, Cytocine)
clusters in coding and non-coding regions
of different organisms \cite{stanley:1995,provata:1997}. Other studies
manifested long range correlations and power laws in the
primary structure of DNA using a variety of statistical methods ranging from
wavelets to linguistic approaches \cite{arneodo,hes2,tsonis:1991,
karlin:1993,herzel:1998,hes:1993}.
\par In recent studies, one of the present authors (AP) and coworkers
have shown that the long range distributions of Pyrine and Pyrimidine clusters
in the non-coding regions of higher eucaryotes 
are related to similar long range distributions 
present at a higher level of genomic organisation:
the level of coding  and non-coding alternating regions
\cite{provata:1999,provata:2002,katsaloulis:2002}. 

\par Non-extensive Statistical Mechanics is particularly fitted to describe
complex structures which present long range correlations, power laws and
fractality \cite{tsallisreview}. In particular, non-extensive statistics have been
used to describe successfully complex spatiotemporal structures in diverse
fields such as high energy physics \cite{bediaga:2000}, turbulence
\cite{turbulence}, 
biological
systems \cite{upadhyaya}, anomalous diffusion \cite{alemany:1994}, 
classical and quantum chaos
\cite{lyra:1998}, interacting particle systems
\cite{latora:2001} and reactive dynamics
\cite{tsekouras:2004}.
\par Classical Statistical Mechanics uses the Boltzmann Gibbs (BG) Entropy, $S_{BG}$, 
defined as:
\begin{eqnarray}
\label{eq01}
\begin{array}{lllllll}
S_{BG} = -\displaystyle {\sum_{i=1}^W} p_i \ln p_i  
\end{array}
\end{eqnarray}
to describe the properties of systems at equilibrium. In Eq. \ref{eq01},
$p_i$ denotes the probabilities of the $i-th$ microscopic state and
the average runs over
the total number of states $W$. This BG entropic form
can not successfully describe systems in which self-organisation,
long range features and scaling are observed. As a generalisation of
Eq. \ref{eq01}, Tsallis and coworkers \cite{tsallis} have introduced 
the non-extensive entropy, defined as:
\begin{eqnarray}
\label{eq02}
S_q  =  \displaystyle {{{1-\displaystyle 
{\sum_{i=1}^{W}}p_i^q}\over{q-1}}}, & {\rm for} & q\ne 1
\end{eqnarray}  
 where $q$ is the non-extensivity exponent. Note that for $q=1$
the classical BG statistics (Eq. \ref{eq01}) is recovered and thus 
departure of the exponent $q$ from the value 1 signals departure
from BG statistics.
\par In relation to non-extensive statistics, 
long range decay maybe obtained by a
non-linear dynamical process expressed by \cite{tsallis:1999}:
\begin{eqnarray}
\label{eq03}
{{d\xi}\over{ds}}=-\kappa_q\xi^q, \>\>\>\>
 {\rm for} \>\>\>\> ( \kappa_q \ge 0, q\ne 1 )
\end{eqnarray}
In particular, for $q > 1$ long range decay is manifested, while
for $q=1$ the well known exponential decay is obtained. The solution
of Eq. \ref{eq03} is:
\begin{eqnarray}
\label{eq04}
\xi (s) &=& \left[ 1-(1-q)\kappa_q (s-1)\right] ^{1/(1-q)},  \>\>\>\>
{\rm for} \>\>\>\> ( \kappa_q \ge 0, q > 1 )\\
\nonumber
&=&  \exp (-\kappa_1 (s-1)), \>\>\>\> {\rm for} \>\>\>\> ( \kappa_1 \ge 0, q=1 )
\end{eqnarray}
with initial condition $\xi (1) =1$.
Thus for $q>1$ a long range law (power law decay) is obtained, 
while for $q=1$ a short range
law (exponential decay) emerges. 
\par For phenomena where two or more dynamical mechanisms act in the system
producing different decay laws in different length (and/or time) scales,
a further phenomenological
generalisation of Eq. \ref{eq03} maybe introduced by addition of terms carrying
different powers \cite{tsallis:1999}. The simplest one carries only
one additional term and is:
\begin{eqnarray}
\label{eq05}
{{d\xi}\over{ds}}=-\kappa_q\xi^q-(\lambda_r-\kappa_q)\xi^r, \>\>\>\>
 {\rm for} \>\>\>\> ( q\le r )
\end{eqnarray}
Note that Eq. \ref{eq03} is recovered for $\kappa_q=\lambda_r\>\> (\forall r)$. 
The solution of Eq. \ref{eq05} can not be written in a simple form
but it may be shown that it consists of two distinct power law
regions, one governed by the exponent $q$ and 
one by the exponent $r$ \cite{tsallis:1999}.
\par
\begin{figure}
\label{fig1}
\includegraphics[width=12.0cm,height=8.0cm]{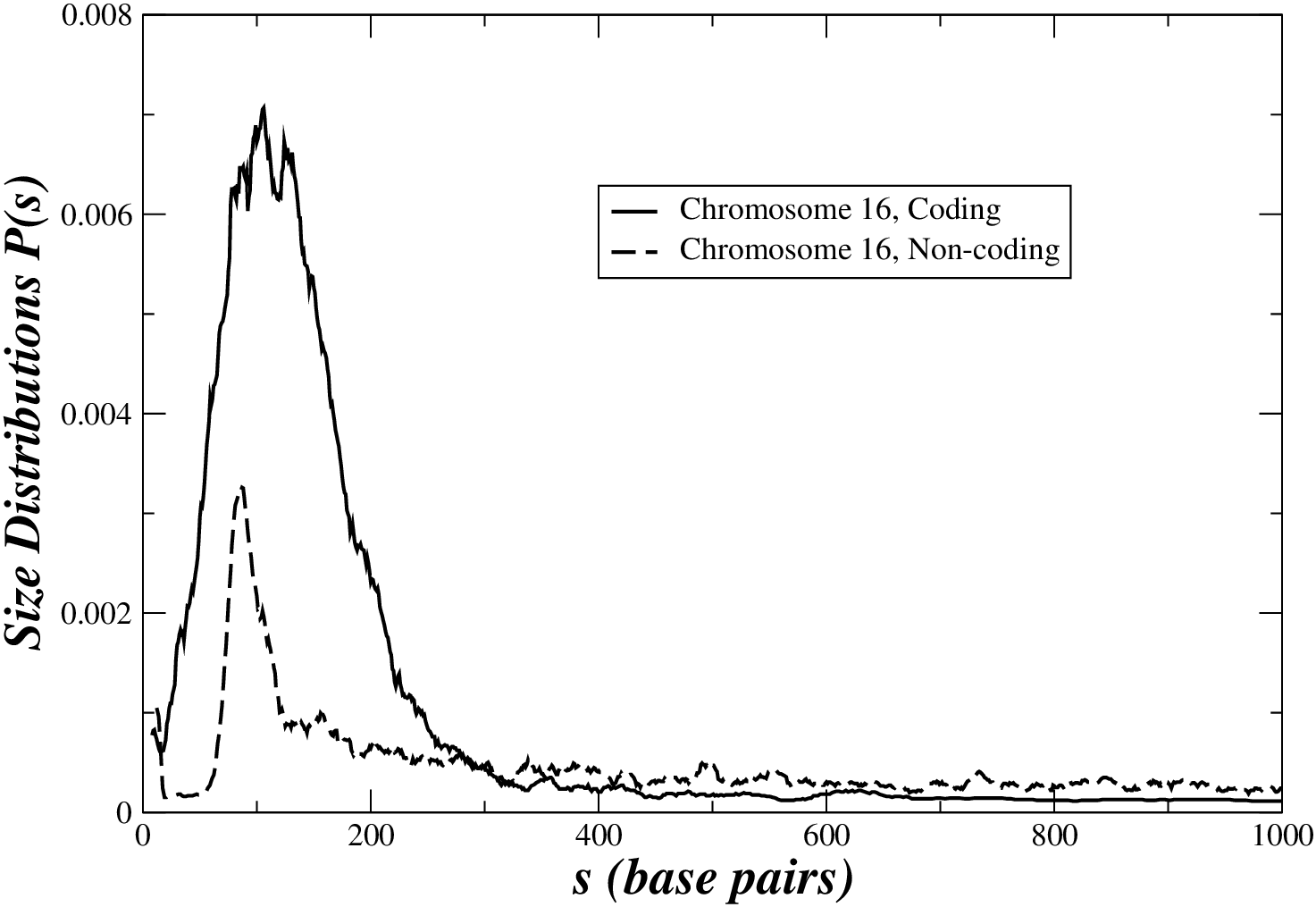}
\caption{Running average over 15 points of 
the size distribution of coding and non-coding DNA in chromosome
16. Only first 1000 bps are shown.}
\end{figure}

\begin{figure}
\includegraphics[width=12.0cm,height=8.0cm]{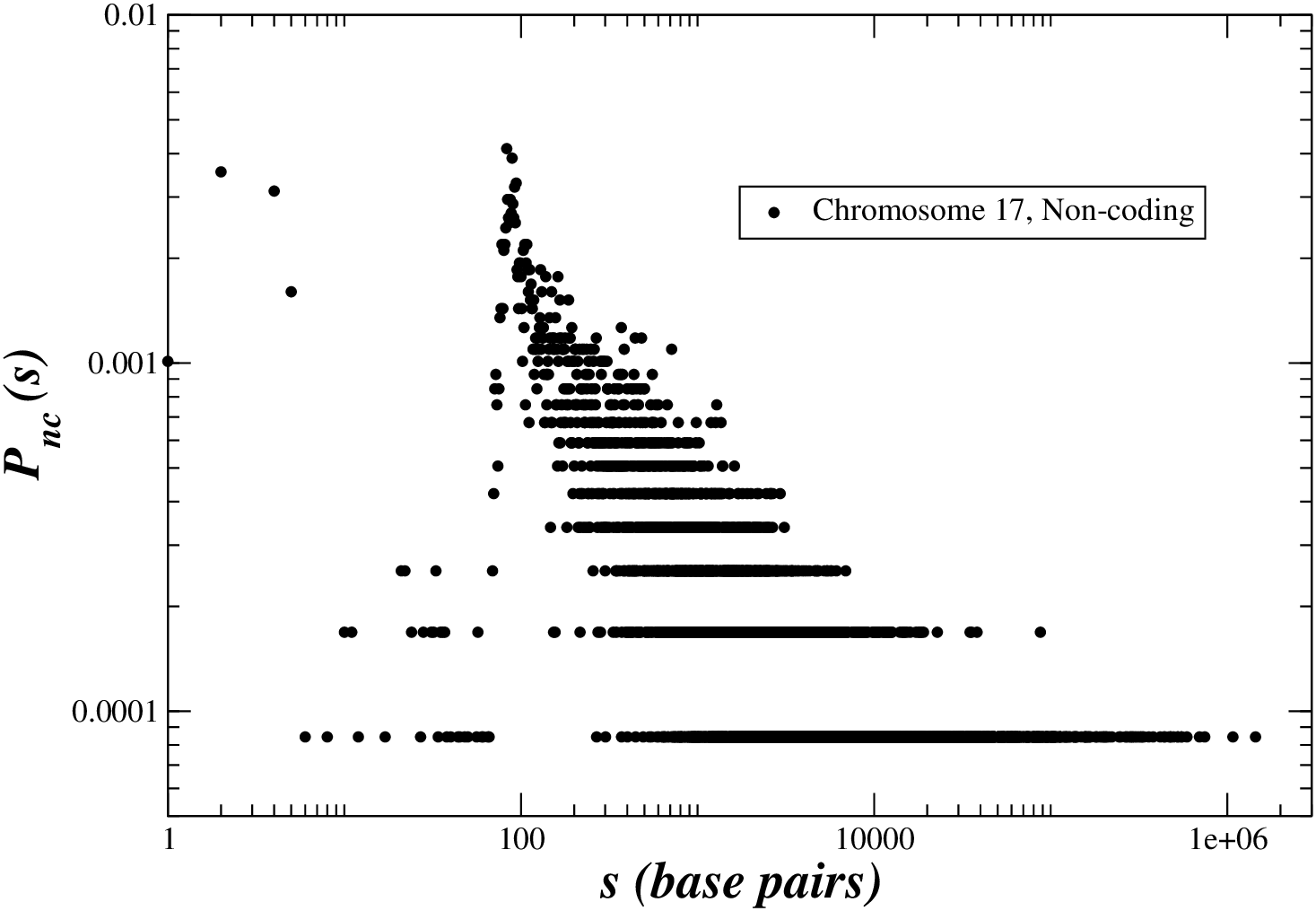}
\caption{ The size distribution of non-coding DNA in chromosome
17 in a double logarithmic scale (all data).}
\label{fig2}
\end{figure}

 In Fig. \ref{fig1} we present the size distribution of coding and
non-coding DNA sequences in chromosome 16. To avoid local fluctuations
running averages are considered over 15 Base Pairs (bps).
For clarity only the first 1000 points are shown. The maximum size of
coding regions is of the order of 7000-8000 bps (reaches 20000bps
for chromosome 19) while the maximum sizes of the non-coding regions 
reach $10^8$ bps. The coding size distributions are rich in small segments
of the order of 100-110 bps and then fall fast, while the non-coding ones
have a similar maximum in the small scales and fall relatively slower.
For comparison we also present the size distribution
of chromosome 17, in Fig. \ref{fig2}, in double logarithmic scale
where the entire $s$-range is shown. 
\par Comparing Figs. \ref{fig1} and \ref{fig2} we note that the
 size distributions of  non-coding DNA, has a complex
form but we may clearly distinguish two regions:
one region at the short scales which is bell-shaped 
and which describes the introns (non-coding regions within genes) and
one region at the larger scales which contains long tail and which describes
the non-coding intergenic regions. 
It is thus natural, at the phenomenological 
level, to use the dynamics of Eq. \ref{eq05} for the description 
of the complex shape of the size distribution of non-coding DNA hoping to
capture these two trends, the introns and the intergenic regions. 
\par

\par In the current study we use non-extensive statistics
to study the size distributions of coding and non-coding
sequences in the human genome which is now near completion. We have selected
to study 4 of the most complete human chromosomes including chromosome 19
which contains the highest percentage of coding.
In the next section we concentrate on the primary
structure of the human genome and we 
give details on the particular data we use.
In sections \ref{sec:cod} and \ref{sec:ncod} we present the analysis of the
size distribution of coding and non-coding DNA, respectively. 
We conclude by summarising our 
main results and discussing some open problems.

\bigskip
\section {The Human Genome}
\label{sec:genome}
\par Although officially the human genome project is announced to be near
completion, in the international EMBL and GenBank genomic data bases
the sequence data deposited varies from 98.91\% 
for chromosome 17 to 43.1\% for chromosome Y. The unknown base pairs
are usually denoted by the letter $N$=(unknown base pair) and they are
either isolated or appear in clusters. The meaning of $N$ is not 
unique. It might denote a base pair which resists to sequencing methods
completely or partially. Resisting partially means that partial information
on the base is known, for example being a Purine or a Pyrimidine.
Another case is that the various laboratories which verify the sequencing 
may not agree on this base pair.
\par In the current project we analyse the complete primary structures
of human
chromosomes 6, 16, 17 for which the $N$ percentage is the smallest and also
chromosome 19, which contains the highest percentage of coding DNA, 3.8\%.
The sequenced percentage presented in the data bases and the coding percentage
of these are shown in Table \ref{table:cod}.
After downloading the chromosomes we isolate the coding and non-coding
segments and calculate their respective size distributions for each of
them. Representative plot is shown in Fig. \ref{fig2}. Due to the heavy fluctuations
in the data we prefer to work with the cumulative distributions
 $\tilde P(s)$  defined as:
\begin{eqnarray}
\tilde P(s)=\int_s^{\infty}P(l)dl
\label{eq06}
\end{eqnarray}
where $P(l)$ is the usual distribution of coding
or non-coding regions of size $l$.  In general, due to summation
 the cumulative
distributions have  better statistical properties than
the usual distribution functions while they keep the main data trends. 
Notice that, if 
the distribution $P(l)$ has the exponential (short range)
form its cumulative $\tilde
P(s)$ will also have the exponential form. If the distribution function
has a power law form of the type:
\begin{eqnarray}
P(l)\sim l^{-1-\mu}
\label{eq061}
\end{eqnarray}
 then the cumulative 
distribution will have a power law form with exponent $-\mu$,
\begin{eqnarray}
\tilde P(s)=\int_s^{\infty}l^{-1-\mu}dl=s^{-\mu}, \;\; 0\le \mu\le 2.
\label{eq07}
\end{eqnarray}
Cumulative diagrams of the four coding and non-coding cumulative
size distributions
are shown in Figs. \ref{fig3} and \ref{fig4}, respectively. The non-extensive
analysis of these distributions follows in the next two sections.

\section{Coding DNA}
\label{sec:cod}
\par
\begin{figure}[ht]
\label{fig3}
\includegraphics[width=6.0cm,height=5.0cm]{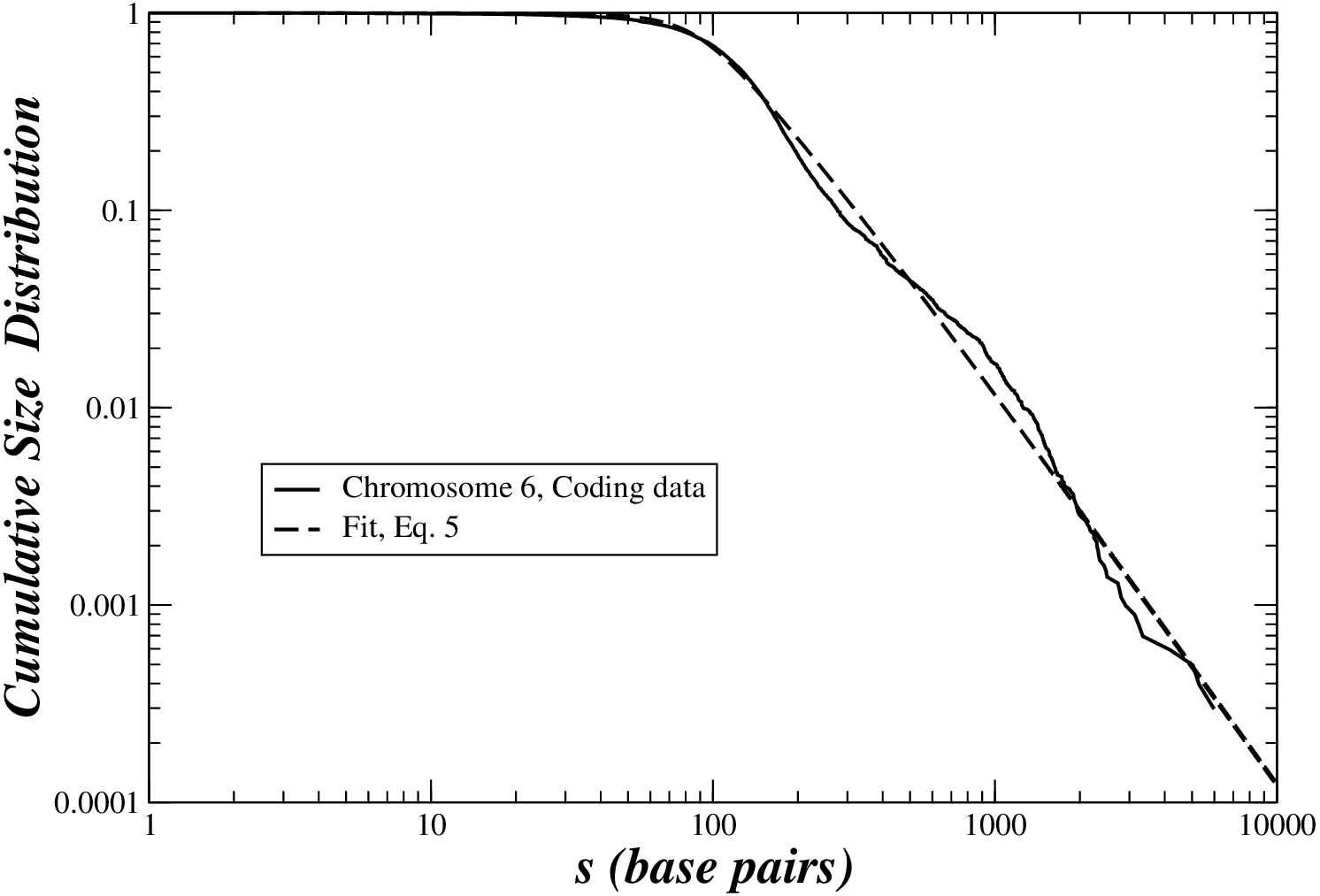}
\includegraphics[width=6.0cm,height=5.0cm]{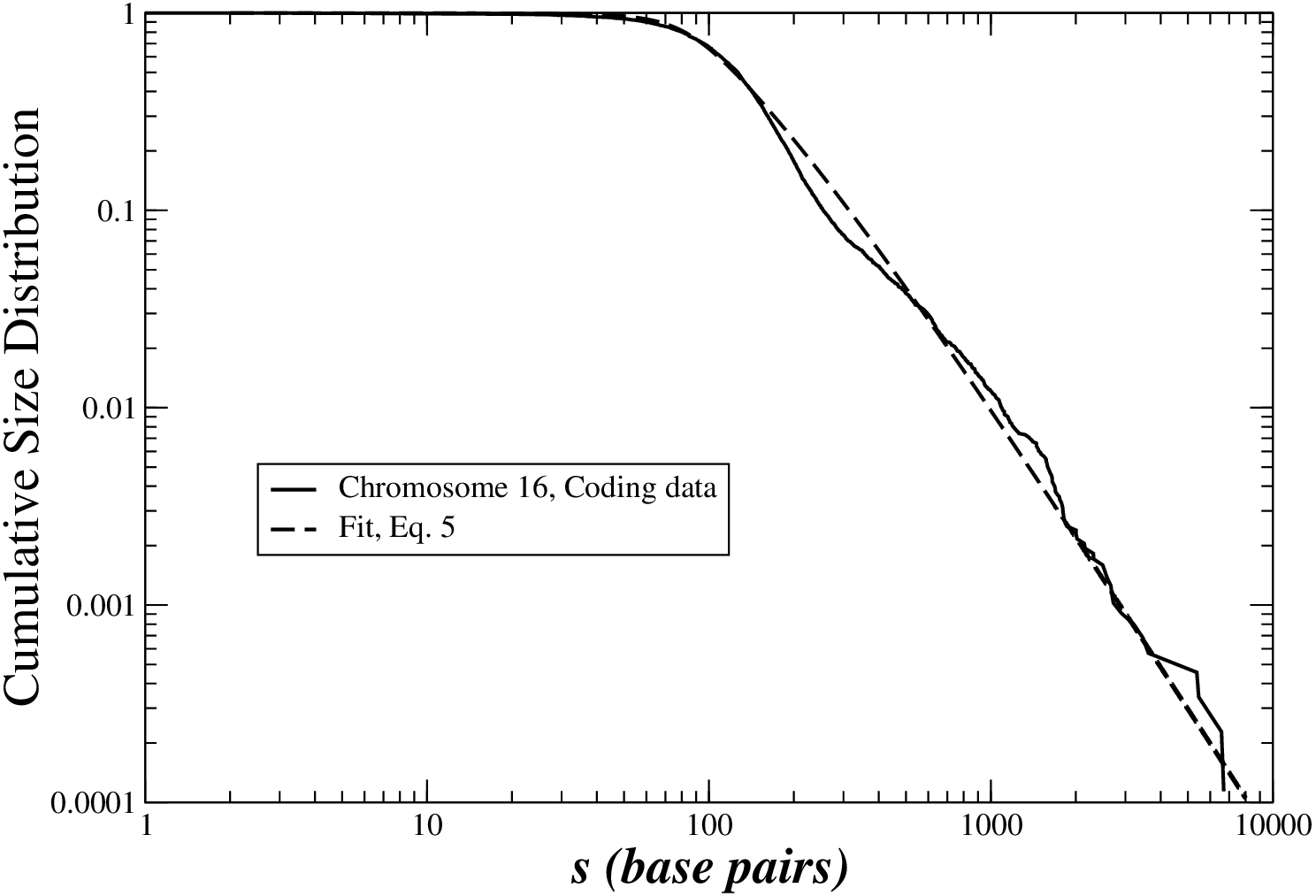}\\ \\ \\
\includegraphics[width=6.0cm,height=5.0cm]{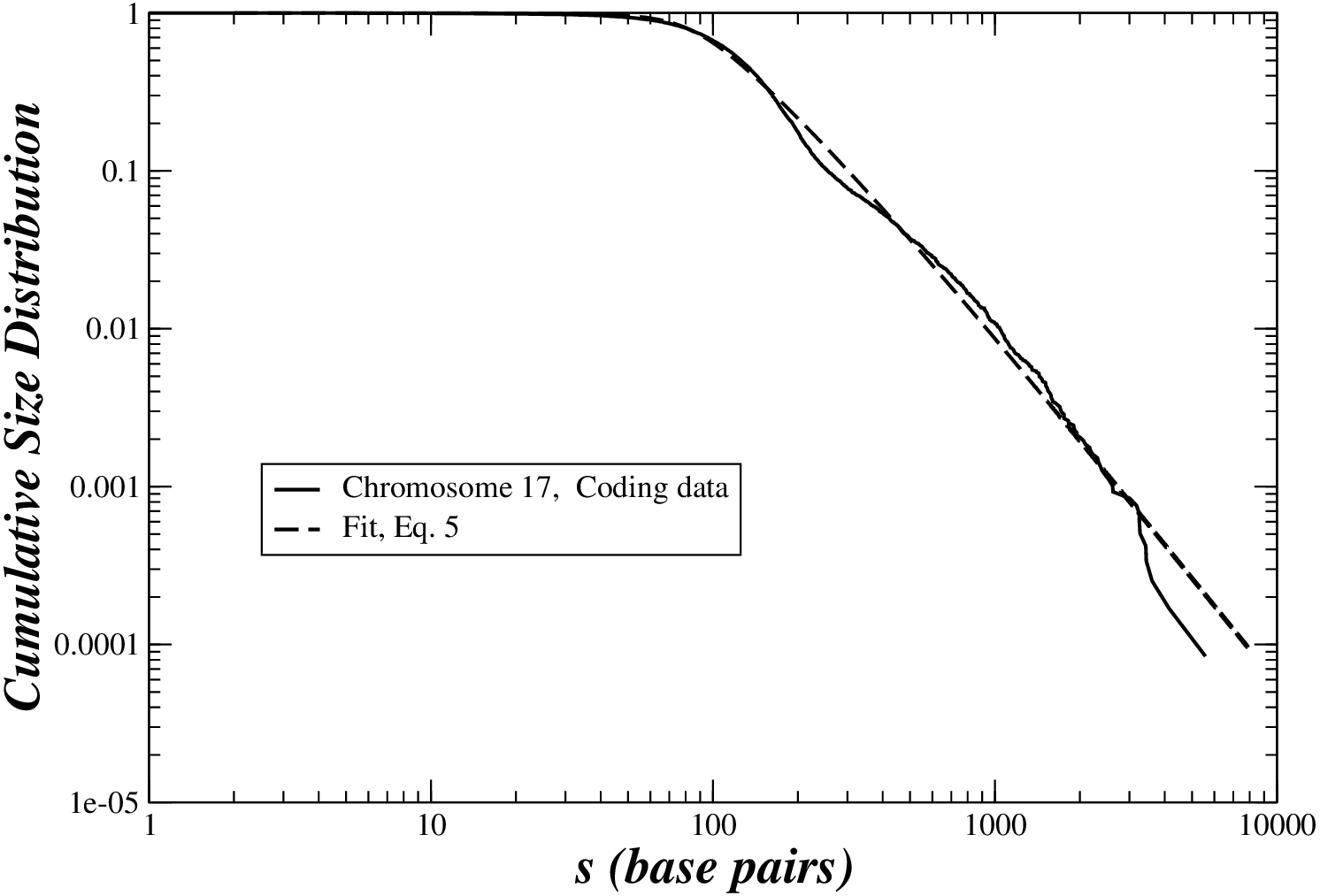}
\includegraphics[width=6.0cm,height=5.0cm]{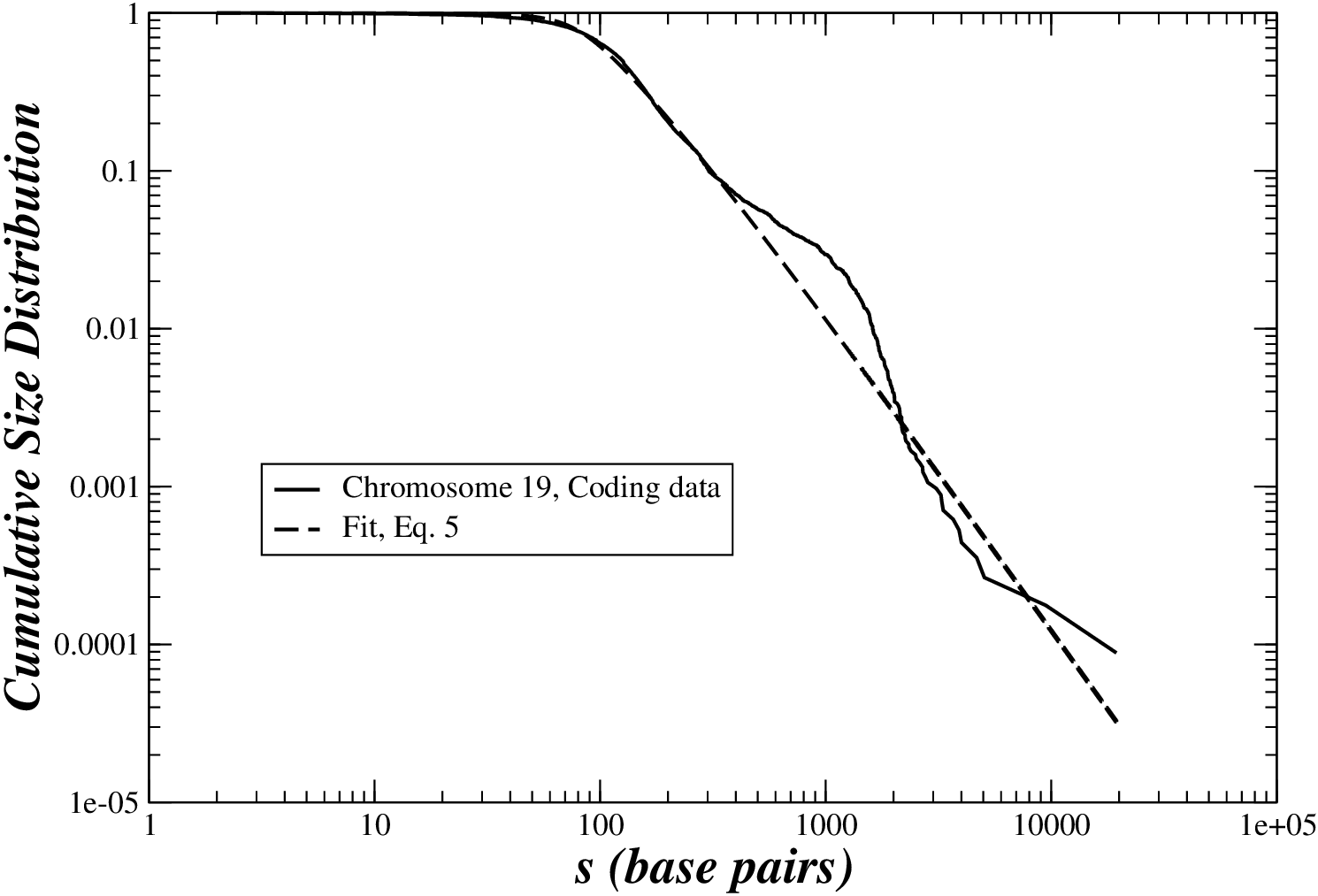}
\caption{ The cumulative size distributions of coding DNA in chromosomes
6, 16, 17 and 19 (solid lines) and the non-linear 
fits using Eq. \ref{eq05} (dashed lines).}
\end{figure}

 As we have already seen in Fig. \ref{fig1} the coding size
distributions have a bell-shape and their tails in the large scales
fall relatively fast. To give a quantitative account for the decay of
the tails of the distributions we plot the 
cumulative size distributions in Fig. \ref{fig3} (solid lines).
\par To describe the shape of the four curves we use the phenomenological 
non-extensive description of Eq. \ref{eq05} and the corresponding
curves are also shown in the same figures (dashed lines). The
theoretical lines approximate well the data. 
The exponents $q$ which 
describe the tails of the distributions vary between $1.45< q <1.50$ for the
four chromosomes and their  specific values are given in Table \ref{table:cod}. 
The non-extensive exponent $q$
corresponds to power law tails of the form Eq. \ref{eq061} with exponent $\mu$
given by 
\begin{eqnarray}
\mu =-1/(1-q).
\label{eq08}
\end{eqnarray}
Thus the tails of the coding size distributions present 
short
range correlations, since $ \mu \ge 2$.   
The exponent $r$ which expresses
the small scale characteristics, takes values between $5.2<r<5.6$ for 
these chromosomes.  Similar results have also been observed 
for the other human chromosomes. The similarity of the two exponents in the four
chromosomes indicate that the same (or similar) dynamical processes have created the
coding parts of all chromosomes during evolution. This dynamics must be of
conservative type in short time scales, since coding DNA changes very slowly
(behaves as an almost-closed system) and this is consistent with short range
correlations \cite{provata:2002}. As the human genome annotation advances,
we expect that the exponents $r$ and $q$ may be modified and/or other exponents
may be needed for a more complete statistical non-extensive description.
\begin{table}
\begin{center}
\begin{tabular} {|c|c|c|c|c|c|c|c|} \hline
Chromo- & Sequenced   &Coding   &$q$  & $r$  & $\kappa_q$  & $\lambda_r$ &$\mu=$ \\  
some& \% & \%  &  & & & & 1/(q-1) \\ \hline
6     & 97.86    &1.03657  &1.50 &  5.2  & 0.018  & 0.00012 & 2.00\\ \hline 
16     & 88.81    &1.67416  &1.45 &  5.7  &  0.017  &0.00009 & 2.22\\ \hline
17     & 98.91    &2.48184  &1.45 &  5.4  &   0.018 &0.00010 & 2.22 \\ \hline
19     & 87.43    &3.39768  &1.50  &  5.6  &  0.018  &0.00012 & 2.20   \\ \hline
\end{tabular}
\caption{ \label{table:cod} Non-extensive exponents and parameters describing the coding
size distributions.}
\end{center}
\end{table}

\section{Non-Coding DNA}
\label{sec:ncod}
The cumulative size distributions of the non-coding DNA in the 
four chromosomes are shown in Fig. \ref{fig4} (solid lines). We observe that the
four distributions have as common characteristic a long tail
which can be expressed in the form of a pure power law \cite{provata:1999}.
In the smaller scales the decay is characterised by a different exponent
which is very similar for the four distributions. 
\par To describe the shape of the four curves we use the phenomenological 
non-extensive description of. Eq. \ref{eq05} and the corresponding
curves are shown in the same figures (dashed lines). The
theoretical lines are very faithful approximations to the data. 
The exponents $q$ which 
describe the long tails of the distributions are very close for the
four chromosomes and their corresponding values are given in 
Table \ref{table:ncod}.
Their values vary between $1.50<q<1.72$. 
The non-extensive exponent $q$
corresponds to a power law of the form Eq. \ref{eq061} with exponent $\mu$
being within the bounds $0\le \mu \le 2$, which indicates clear long
range correlations. In the case of chromosome 19, which (up to now) contains
the highest coding percentage amongst all human chromosomes, the value of $\mu$
calculated through Eq. {\ref{eq08} is equal to 2, which is border line case
between short and long range correlations. On the other hand, 
the exponent $r$ which expresses
the small scale characteristics, takes values between $2.4<r<3.2$ for 
these chromosomes. Similar results have also been observed 
for all other human chromosomes.
\begin{figure}[ht]
\label{fig4}
\includegraphics[width=6.0cm,height=5.0cm]{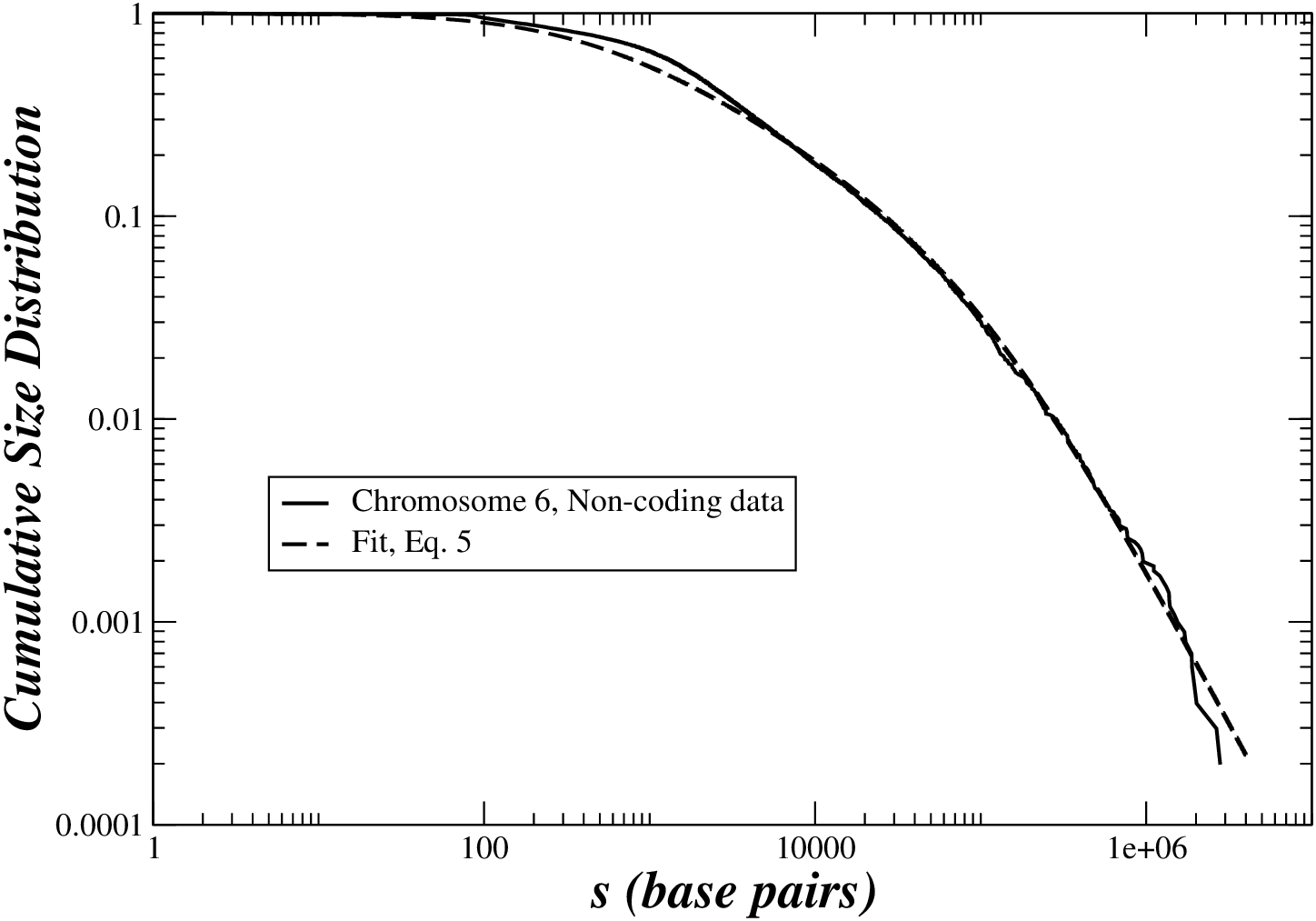}
\includegraphics[width=6.0cm,height=5.0cm]{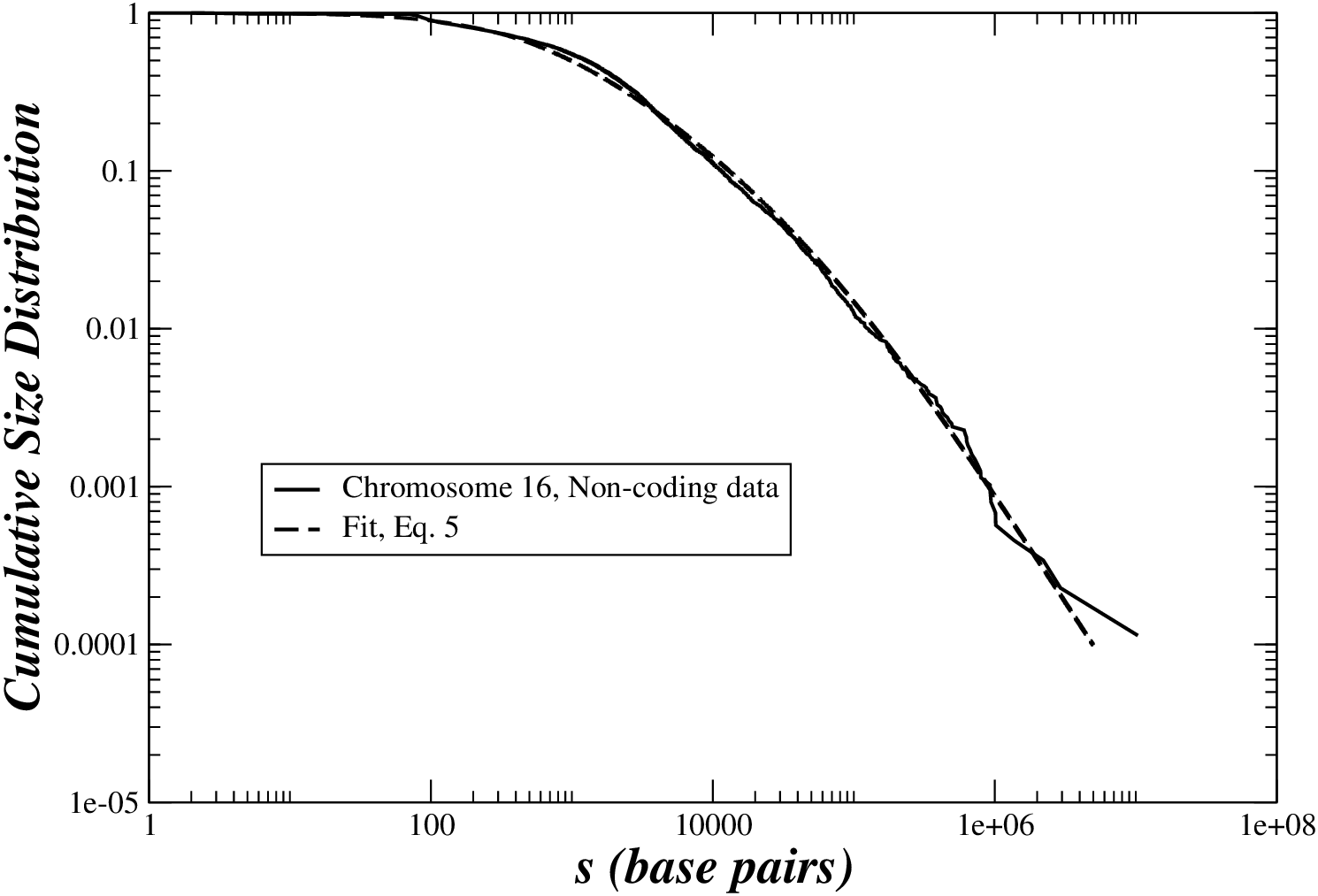}\\ \\

\includegraphics[width=6.0cm,height=5.0cm]{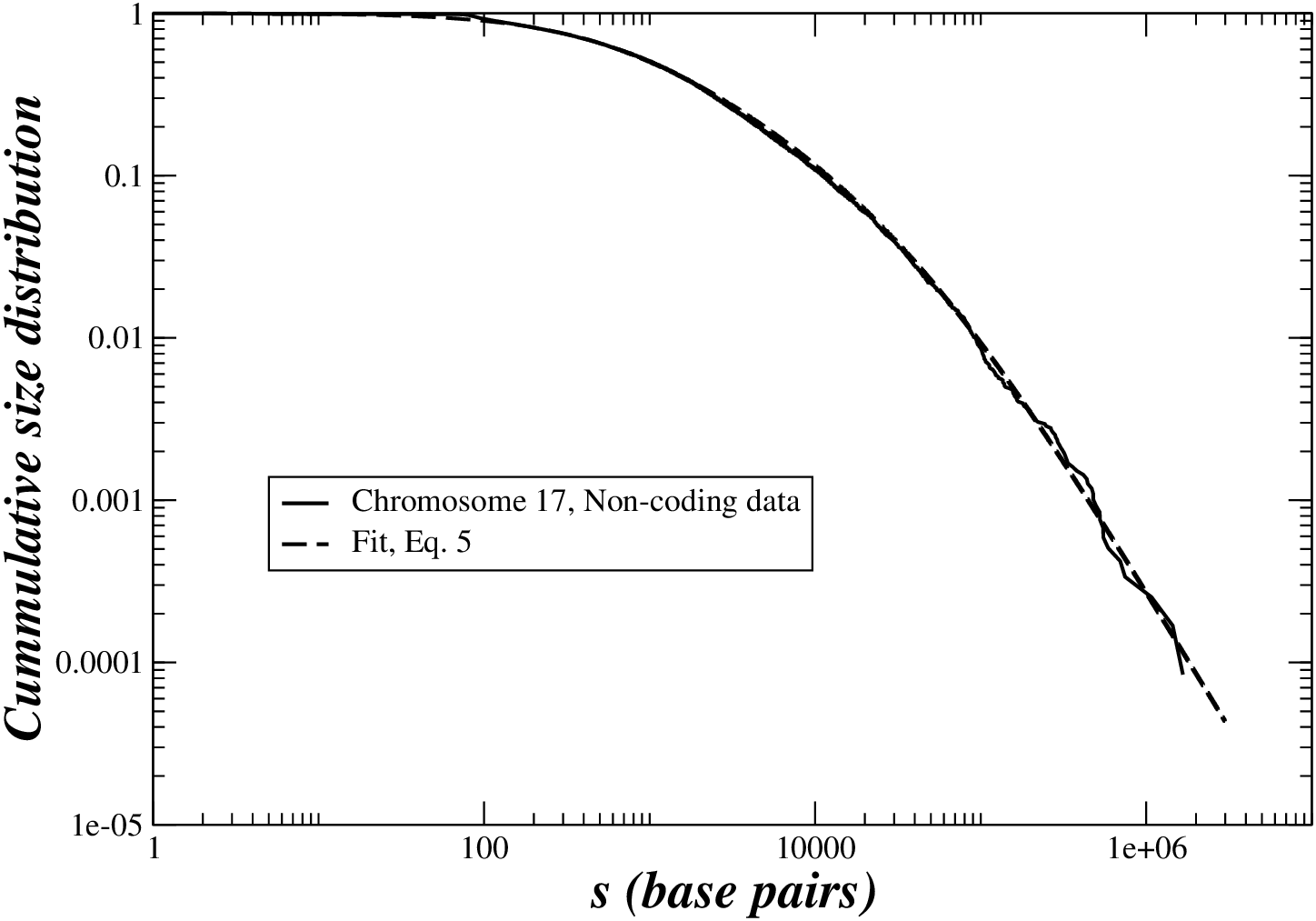}
\includegraphics[width=6.0cm,height=5.0cm]{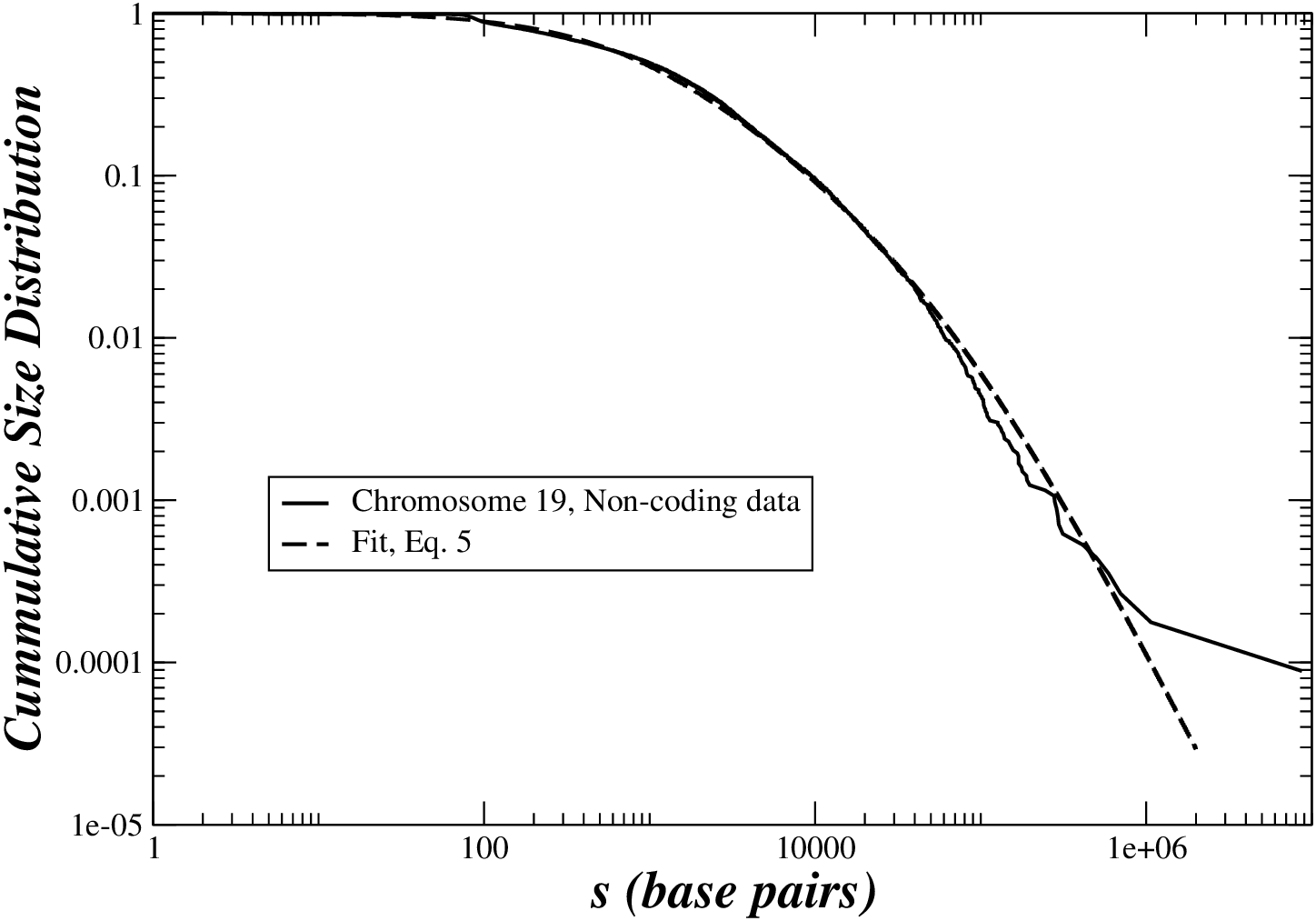}
\caption{ The cumulative size distributions of non-coding DNA in chromosomes
6, 16, 17 and 19 (solid lines) and the non-linear fits
 using Eq. \ref{eq05} (dashed lines).}
\end{figure}

\par The different small and large scale behaviors observed in the
size distribution of the non-coding indicates that different dynamical mechanisms are
involved in the formation of small non-coding segments (which are usually
found as introns in the genes) and in the large non-coding areas, or intergenic 
regions which are found between genes and between families of genes.
The intergenic regions are extended non-coding regions which can support 
extensive (massive) influx and outflux of genomic material. Thus the 
ensemble of intergenic regions acts as an open system which supports
exchange with the environment. In open systems, out of equilibrium, 
power laws and  long range correlations which may be regarded
 as expression of non-extensive
or edge of chaos dynamics, naturally emerge. Open aggregating systems,
with influx mechanisms similar to the ones involved in
genomic evolution and which lead to long range correlations
 are presented in reference \cite{provata:2002}. On the other hand,
the non-coding segments found within genes, called also introns are
less supportive to external influences because often they include
functional strings. Thus they behave more like closed systems and
thus the dynamics must be very different, which is also expressed
by the difference in the exponents $q$ and $r$.
\begin{table}
\begin{center}
\begin{tabular} {|c|c|c|c|c|c|} \hline
Chromo-   &$q$  & $r$  & $\kappa_q$  & $\lambda_r$ & $\mu$= \\  
some&   & & & & $1/(q-1)$ \\ \hline
6      &1.65 &  3.2  & 0.00009  & 0.00120   &1.54\\ \hline 
16       &1.72 &  2.7  &  0.00021  &0.00124   &1.39\\ \hline
17       &1.59 &  2.7  &   0.00021 &0.00118   &1.69 \\ \hline
19       &1.50  &  2.4  &  0.00018  &0.00124   &2.00   \\ \hline
\end{tabular}
\end{center}
\caption{\label{table:ncod} Non-extensive exponents and parameters describing the non-coding
size distributions.}
\end{table}

\section{ Conclusions}
\label{sec:concl}
\bigskip
\par 
We have studied the size distribution of all known coding and non-coding
sequences in human chromosomes 6, 16, 17 and 19. The first three were selected 
as representatives of the most completely sequenced chromosomes while 
chromosome 19 has the highest, up to date, coding percentage. We have found that
the decay of the non-coding size distributions is consistent
 with non-extensive dynamics as expressed by non-linear Eq. \ref{eq05}.
Moreover, we have shown that the non-coding presents two distinct regions,
one large scale region,
related to the intergenic non-coding DNA which presents a decaying exponent
$1.5 \le q \le 1.72$, and a second short scale region related to the introns
(non-coding DNA within genes) which presents a decaying exponent $r > 2.4$.
On the contrary, short range correlations have been observed in the tails
of the coding distributions with non-extensive exponents 
$1.45 \le q \le 1.50$.
This is consistent with earlier observed
long (short) range correlations in the non-coding (coding)
size distributions of higher eucaryotes.
All other human chromosomes demonstrate similar characteristics. 
\par A more
detailed analysis could involve the use of more terms with different
exponents in Eq. \ref{eq05}, in order to capture more details such as
the dynamical exponent which govern non-coding distances between  
families of homologous genes (they may be governed by one of the current exponents,
$q$ or $r$, or by a third one). 
\par It is true that today
the human chromosomes may be close to full sequencing
 but their complete annotation will take much longer. 
This means that there are still  coding
sequences which are not discovered within the genome. Thus we expect that
with the advancement of DNA annotation, which is the next major step
in genomics after sequencing, we will be able to give more precise, final
values to the exponents $q$ and $r$ for the human genome. Also the
study in parallel of
 the genomes of other organisms, as they become sequenced and annotated,
will allow for a comparative analysis of genomic data between different
classes of
organisms.

{\bf Acknowledgments}\\
The authors would like to thank Prof. C. Tsallis for suggesting this approach
and Prof. K. Trougkos for helpful discussions.

\end{document}